\documentclass[fleqn,twocolumn]{elsart3p}
\usepackage{graphics}
\usepackage{graphicx}
\usepackage{epsfig}
\usepackage{amssymb}

\begin{document}

\begin{frontmatter}

\title{The primary cosmic-ray energy spectrum measured with the Tunka-133 array
}
\author[b]{N.M.~Budnev},
\author[d]{A.~Chiavassa}, 
\author[b]{O.A.~Gress},
\author[b]{T.I.~Gress},
\author[b]{A.N.~Dyachok},
\author[a]{N.I.~Karpov},
\author[a]{N.N.~Kalmykov},
\author[a]{E.E.~Korosteleva},
\author[a]{V.A.~Kozhin},
\author[a]{L.A.~Kuzmichev},
\author[c]{B.K.~Lubsandorzhiev},
\author[a]{N.B.~Lubsandorzhiev},
\author[b]{R.R.~Mirgazov},
\author[a]{E.A.~Osipova},
\author[a]{M.I.~Panasyuk},
\author[b]{L.V.~Pankov},
\author[a]{E.G.~Popova},
\author[a]{V.V.~Prosin\corauthref{cor}},
\corauth[cor]{Corresponding author.}
\ead{v-prosin@yandex.ru}
\author[g]{V.S.~Ptuskin},
\author[b]{Yu.A.~Semeney},
\author[a]{A.A.~Silaev},
\author[a]{A.A.~Silaev(junior)},
\author[a]{A.V.~Skurikhin},
\author[e]{C.~Spiering},
\author[a]{L.G.~Sveshnikova},
\author[b]{A.V.~Zagorodnikov},

\address[a]{Skobeltsyn Institute of Nuclear Physics MSU, Moscow, Russian Federation}
\address[b]{Irkutsk State University, Irkutsk, Russian Federation}
\address[c]{Institute for Nuclear Research of the Russian Academy of Sciences,
Russian Federation}
\address[d]{Dipartimento di Fisica  dell'Universita and INFN,
Torino, Italy}
\address[e]{DESY, Zeuthen, Germany}
\address[g]{IZMIRAN, Troitsk, Moscow Region, Russia}



\begin{abstract}
The EAS Cherenkov light array Tunka-133, with $\sim$ 3 km$^2$ geometric area, is 
taking data since 2009.The array permits a detailed study of 
energy spectrum and mass composition of cosmic rays in the energy range from 
$6\cdot 10^{15}$ to $10^{18}$\,eV.
We describe the methods of time and amplitude calibration of the array and the methods
of EAS parameters reconstruction. We present the all-particle energy spectrum, based 
on 7 seasons of operation.

\vspace{1pc}
\end{abstract}
\end{frontmatter}


\section{Introduction}
\label{s1}
The study of energy spectrum and mass composition of primary cosmic rays  in the energy range of
$10^{15}$ - $10^{18}$\,eV is of crucial importance for understanding origin
and propagation of cosmic rays (CR) in the Galaxy. An increasing dominance of heavy 
nuclei from
the "knee" up to $10^{17}$\,eV  indicates the energy limit of CR acceleration in the standard model
of particle acceleration in supernova remnants, see review \cite{bell}.
At higher energies, the composition becomes lighter again up to $2\cdot 10^{18}$\,eV. 
This may indicate a transition to an extragalactic origin of CR or the dominance of 
new Galactic sources of unknown nature, see \cite{blasi} for discussion.

To measure the primary energy spectrum and mass composition of cosmic rays
in the energy range mentioned above,  
the Tunka-133 array \cite{Tunka-133_2012,Tunka-133_2011} 
with nearly 3 km$^2$ geometrical
area has been deployed in the Tunka Valley, Siberia. It
records the Cherenkov light from Extensive Air Showers (EAS) using the  Earth atmosphere as a huge
calorimeter.

\section{The Tunka-133 Array}

\begin{figure}[ht]
\includegraphics[width=75mm]{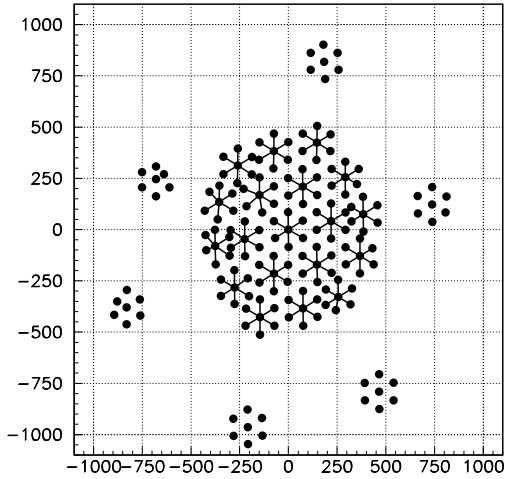} 
\caption{Layout of the Tunka-133 array.}
\label{fig:1}
\end{figure}

The Tunka-133 array is located at an altitude of $670$~m a.s.l.
It consists of 175 optical detectors.
The detectors are grouped into 25 clusters with seven detectors each -- six
hexagonally arranged detectors and one in the center. The distance between the
detectors in a cluster is 85\,m (see Fig.\,\ref{fig:1}). 19 of the clusters are arranged as a dense 
central part of the array with a radius of about 500 m. These $19 \times 7 = 133$ stations
gave the array its name. The remaining  clusters are installed at 
distances of $\approx 1$\,km around the central part.

Each optical detector consists of a metallic cylinder with 50\,cm diameter, containing a
single PMT with a hemispherical photocathode of \mbox{20 cm}. The container 
window consists of plexiglass and points vertically upward.
It is heated against hoar-frost and dew. The detector is equipped with a remotely
controlled lid protecting the PMT from daylight and precipitation. 

\begin{figure}[ht]
\includegraphics[width=75mm]{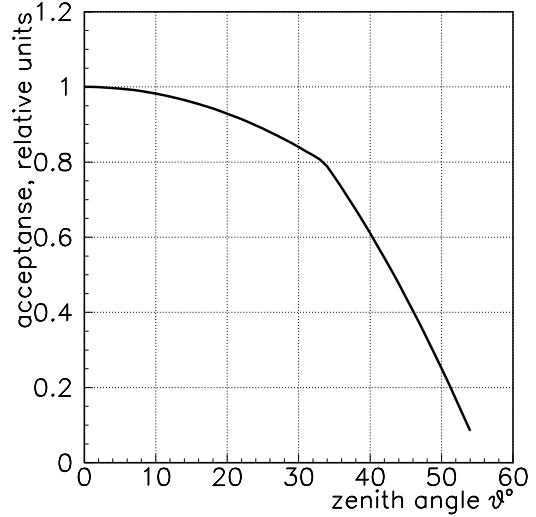} 
\caption{Acceptance of the optical detector vs. zenith angle.}
\label{fig:2}
\end{figure}

\begin{figure}[ht]
\includegraphics[width=75mm]{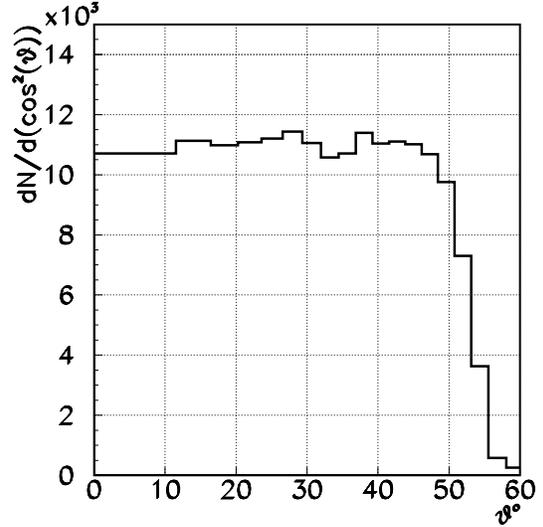} 
\caption{Zenith angle distribution of showers with a reconstructed energy larger
than 15.0\,PeV.}
\label{fig:3}
\end{figure}

Figure \ref{fig:2} shows the detector acceptance as a function of the zenith angle. 
The curve reflects the
influence of the window edge as well as the atmosphere absorption due to 
Rayleigh and aerosol scattering of the Cherenkov light.  
To check the simulated detector acceptance we analyse the zenith angle
distribution of the recorded EAS with reconstructed energies larger than $10.0$\,PeV.
%
%
The resulting zenith angle distribution 
is shown in Fig.\,\ref{fig:3}. The 
distribution is flat up to 50$^{\circ}$.

The PMT pulses are amplified by pre-amplifiers and sent 
via 95\,m coaxial cables RG58
to the electronics box in the center of each cluster and digitized there. 
The digital information is stored in the cycle buffer memory. The memory is stopped,
and the data is sent to the central DAQ system in case of a cluster trigger appearence. 
The trigger condition is the appearence of pulses exceeding the 
digital threshold (see apparatus description \cite{t133_2005}) at 3 or more detectors 
of a cluster within a time gate of 0.5 $\mu$s.

The minimum pulse width is about 20\,ns. The dynamic range of the
amplitude measurement is about $3\cdot 10^4$. This is achieved by means of
two channels for each detector extracting the signals from the anode and an
intermediate dynode of the PMT with different additional amplification factors. 

The data of all clusters is collected by the central DAQ and transferred via
fiber-optics cables.

The amplitude calibration of the detectors was carried out in two
steps similar to that used at the previous experiment Tunka-25. 
The details are described in \cite{t25}.

\section{Calibration}

\subsection{Timing calibration}

The timing calibration of the detectors is performed using  the experimental data from the 
recorded air showers.
 
The calibration inside a single cluster consists of 
a multiple reconstruction of zenith angle $\theta$ and azimuth angle $\phi$ for all 
the recorded showers, with an analysis of the residual difference between the 
measured 
and the theoretically expected delay for 6 peripheral detectors of a cluster. 
For calibration purposes at this step we use a plane model of the shower front. 
Modelling a curved shower front demands knowledge of the 
EAS core position which we reconstruct only at later steps. 
The difference between  plane front and a realistic curved front leads to a larger dispersion  
of the residuals $\Delta T_{i}$  but it does not distort very much the 
mean value of $\Delta T_i $ ($\langle\Delta T_i\rangle$) because of an almost uniform core distribution 
across a cluster area and the limited size of the clusters.

The residual $\langle\Delta T_i\rangle$ is treated as an additional apparatus shift and  
added to the calibration shift. Then the procedure of the arrival direction 
reconstruction is repeated several times till all the detector shifts become less 
than  $\langle\Delta T_i\rangle = 1$\,ns. 

The obtained  corrections of time shifts are used for a preliminary reconstruction
of the EAS core position with a plane model of the shower front, using that single 
cluster which recorded the largest amplitudes of pulses.

At the second step of the timing calibration we select large showers with more
than 8 hit clusters and reconstruct the shower 
front assuming a curved shape derived from CORSIKA simulations and described in 
the next section.

The space basement of optical stations for such events is more than 500\,m. 
Therefore the error of cluster synchronization (about 10\,ns) leads to  
a maximum error in the arrival direction of $0.4^{\circ}$.

Finally the two calibration parameters are {\it a)} the delays between the central 
stations of the different clusters and {\it b)} the delays of the peripheral 
detectors within a cluster relatively to their central one.

The corrections of the central station delays are obtained applying the multi-cluster 
procedure for the central stations only  to a single cluster, one but with 
a curved shape of the shower front.

\begin{figure}[ht]
\includegraphics[width=75mm]{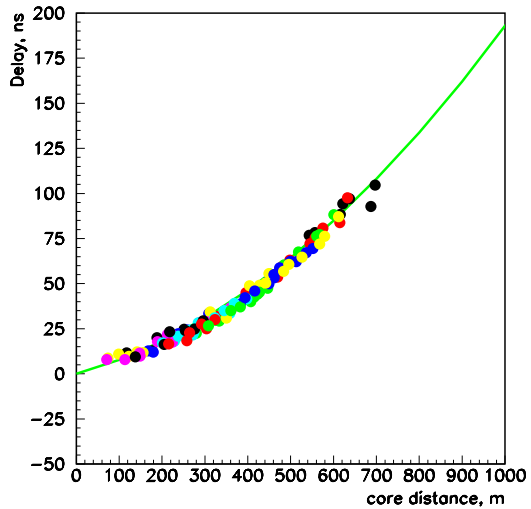} 
\caption{An EAS front reconstructed from an experimental event, $E_0 = 6.8\times 10^{17}$ eV,
zenith angle  $\theta$ = 19.2$^{\circ}$}
\label{fig:4}
\end{figure}

The last correction of the time shifts is made
synchronously for all 6 peripheral detectors of a cluster rotating their common 
plane so that the EAS axis direction measured by each single cluster coincides with 
the direction derived
from a multi-cluster reconstruction with an accuracy better than $0.5^{\circ}$.
An example of an EAS front reconstructed after the time calibration 
procedure is presented in Fig.\,\ref{fig:4}.

\subsection{Relative amplitude calibration}

The amplitude calibration of the detectors was carried out in two
steps \cite{t25}. The first step is a relative calibration, i.e. a derivation of the 
amplitude alignment coefficients for all 
individual  detector outputs when the optical detectors
were illuminated by the same light flashes. 

Differences in gain and 
quantum efficiency of individual detectors resulted in 
output signal amplitude variations of a factor 2-3. 

As the apertures of all detectors are the same, these coefficients were
determined by comparing the amplitude spectra of the EAS Cherenkov light flashes.
The residual dispersion after the renormalizing was less than 10\%.
Due to the stability of the EAS flux, 
the variation of the coefficient  
essentially reflects the time drift of the PMT gain as well as 
the relative atmospheric transparency.

The second step was the absolute calibration, made by
normalization of the obtained integral
energy spectrum to a reference spectrum.
The details of this procedure will be described in section 4.5.

\section{EAS parameter reconstruction}

\subsection{Fitting the Cherenkov light pulse}

\begin{figure}[ht]
  \includegraphics[width=75mm]{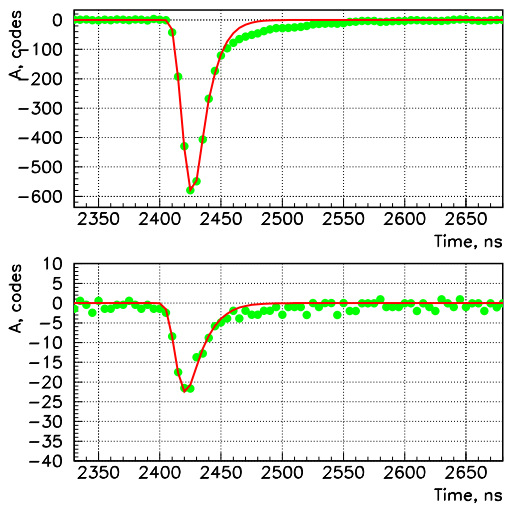}
  \caption{Experimentally recorded pulses. The upper panel displays a pulse from the high gain channel 
(anode), the lower panel displays the same pulse recorded with a low gain channel (dynode). Each curve 
  is a fit of the experimental points by  expression (1).}
  \label{fig:5}
 \end{figure}

The raw data record for each Cherenkov light
detector contains 1024 amplitude values $a_k$ in steps of 5\,ns 
(Fig.\,\ref{fig:5}).
Thus each pulse waveform is recorded over 5\,$\mu$s.

The waveform of the light pulse is too complicated
to be fitted with any simple function like a Gaussian or gamma function.
We have constructed a function which separately approximates front and  tail of 
a pulse \cite{0492}. The expression is based on 4 independent
variables: the pulse amplitude $A$, the time of the pulse maximum $t_{max}$,
and $t_{front}$ and $t_{tail}$  -- the times of the front (pulse reaching 0.1\,$A$) 
and the tail (pulse falling to 0.1\,$A$).

From these variables we derive:\\
$$
\begin{array}{lc}
x = t-t_{max}\\
f = |x/t_{front}|\\
g = x/t_{tail}\\
h=\left\{
\begin{array}{lc}
\!\ 1.7-0.5\cdot g, & \mbox{if}\hspace{5mm} g < 0.8\\
\!\ 1.3, & \mbox{if}\hspace{5mm} g\geq 0.8\\
\end{array} \right.
\end{array}
$$

Using these variables the function for fitting the pulse waveform is as follows:

\begin{equation}
\label{math41}
 a(t)=\left\{
\begin{array}{lc}
\!\ A\cdot \exp( -f^{2+0.5\cdot f}), & \mbox{if}\hspace{5mm} x\leq 0 \\
\!\ A\cdot \exp( -g^h), & \mbox{if}\hspace{5mm} x>0 
\end{array} \right. 
\end{equation}

This expression reasonably well approximates the pulse waveform from the level of 
0.1$A$ at
the front edge to the level of 0.2$A$ at the droop of the pulse. This time range 
is enough
to determine by fitting function (1) such pulse parameters as peak 
amplitude $A_i$ and 
front delay $t_i$ at the level of  $0.25\cdot A_i$.
The pulse area $Q_i$ is measured by digital integration of the waveform 
starting from the level $0.05\cdot A_i$ at the front till the same level at 
the tail. 
A fourth pulse parameter 
is the effective width  $\tau_{eff} = Q_i/(1.24\cdot A_i)$.
The accuracy of this parameter is better than that of the pulse width (FWHM) 
used in earlier reports. 

An example of a fit of experimental pulses with expression (1) 
is shown in  Fig.\,\ref{fig:5}.

\subsection{Arrival direction}

Zenith angle $\theta$ and azimuth angle $\phi$  of the shower direction are reconstructed by 
fitting the measured delays with the curve shower front:
$\Delta T = T_i - T_f = R\cdot (R + 500) / (c\cdot F)$, where $T_f$ is the estimated delay 
for a plane front, 
$R$ the perpendicular distance from the shower axis in meters, $c$ the 
speed of 
light and $F$ the third variable parameter (together with $\theta$ and $\phi$). 
This approximation is derived from the 
analysis of simulated showers with CORSIKA. The formula, on the one hand, has a 
non-zero value of the derivative at $R = 0$ (conical shape, typical for 
Cherenkov 
radiation at short distances from the axis) and on the other hand, has only 
one shape parameter which is essential for processing of relatively small 
showers. To obtain $R$ one needs to know the EAS core position.

\subsection{Core position}

\begin{figure}[ht]
\includegraphics[width=75mm]{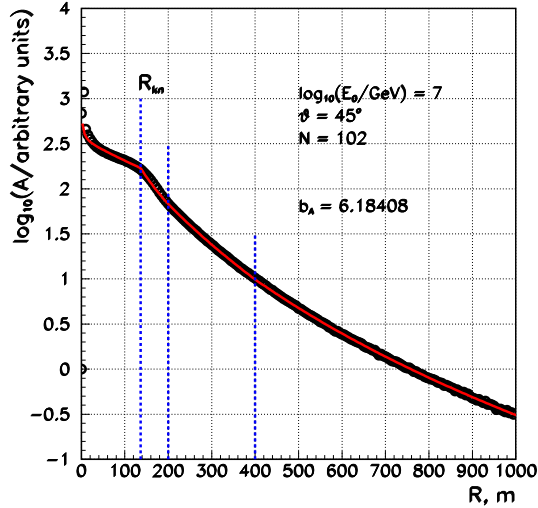} 
\caption{An example of the ADF of a simulated event fitted with expression 3.}
\label{fig:6}
\end{figure}

\begin{figure}[ht]
\includegraphics[width=75mm]{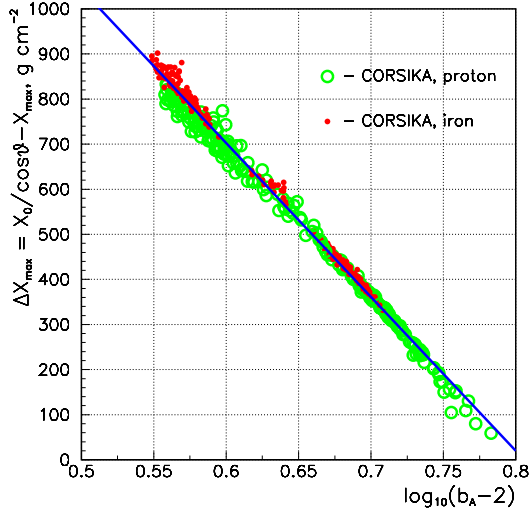} 
\caption{Correlation between $b_A$ and the thickness of matter between observational level 
and the EAS maximum depth.}
\label{fig:7}
\end{figure}

\begin{figure}[ht]
\includegraphics[width=75mm]{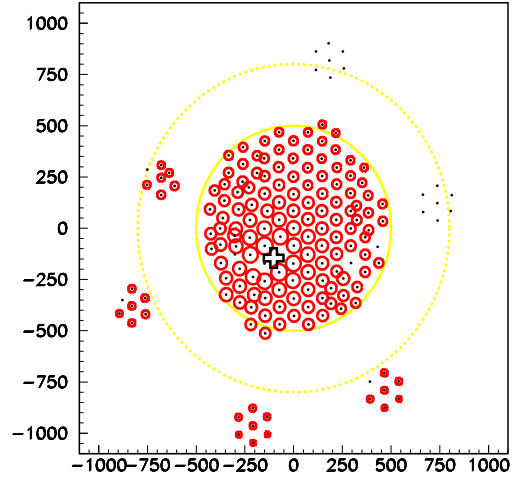} 
\caption{Example of an experimental EAS core reconstruction. The radius of
each station circle is proportional to  $\log Q_i$. The cross marks the 
reconstructed position of the shower core.}
\label{fig:8}
\end{figure} 

\begin{figure}[ht]
\includegraphics[width=75mm]{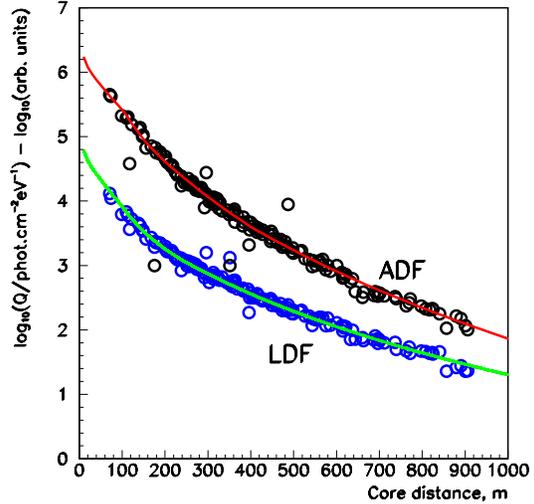} 
\caption{Pulse amplitudes of the event shown in Fig.\,\ref{fig:6}. 
Upper curve: fitted with ADF ( expression 3, arbitary units).
Lower curve: fitted with LDF (expression 6).}
\label{fig:9}
\end{figure} 

The reconstruction of the EAS core position is performed by  
fitting the measured amplitudes $A_i$ with an 
amplitude distance function (ADF): 

\begin{equation}
A(R)=A(200)\cdot p(R),
\end{equation}

The shape of the ADF was been studied with CORSIKA simulated events. It has been checked for 
simulated events in the energy range $10^{15} - 10^{18}$\,eV, for primary protons and iron 
nuclei, and for zenith angles from $0^{\circ}$ to $45^{\circ}$. 
It cannot be fitted with any simple function.
Therefore 
we have desinged a function consiting of four different parametrizations for four different 
distance ranges (see \cite{l7}). 

The first one closest to the core
is an exponential function with variable parameter $R_0$ and addition of a pole close to $R = 0$\,m. 
This parametrization is changed to a
power law function at the variable distance marked as "knee" ($R_{\mbox{kn}}$). The index of the power 
law $b$ is also variable. This function is changed to a Linsley function with a variable 
parameter $a$ at the distance range 200 - 400 m. The last parametrization is the same function with
fixed value of $a = 1$.     
The function $p(R)$ is a fit to four different parametrizations with respect to the
distance $R$ to the shower core (in meters):

\begin{equation}
 p(R)=\left\{
\begin{array}{lc}
\!\exp\left(\frac{(R_{\mbox{kn}}-R)}{R_0}\left(1+\frac{3}{R+2}\right)\right),\\
\hspace{25mm} R<R_{\mbox{kn}} \\
\!\left(\frac{200}{R}\right)^{b},\hspace{7mm} R_{\mbox{kn}}\leq R\leq
200\,\mbox{m} \\
\!\left(\left(\frac{R}{200}+a\right)/{(1+a)}\right)^{-b_A},\\
\hspace{25mm} 200\,\mbox{m}\leq R\leq 400\,\mbox{m}\\
\!\left(\left(\frac{R}{200}+1\right)/{2}\right)^{-b_A},\\
\hspace{25mm} R>400\,\mbox{m}\\
\end{array} \right.
\end{equation}

All four variables in equation (3) ($R_0$, $R_{\mbox{kn}}$,
$a$ and $b$), describing the ADF shape in  different ranges of core distance $R$ 
are related to a single parameter of the ADF shape --
the steepness $b_A$:

\begin{equation}
\begin{array}{lc}
d = b_A-5\\ 
D = \log_{10}(d),\hspace{4mm} \\
R_0 = 275/d^{2},\hspace{4mm} \\
R_{\mbox{kn}} = 145 - 115\cdot D,\hspace{14mm} \\
a = 0.89 - 0.29\cdot D,\hspace{14mm} \\
b =\left\{
\begin{array}{lr}
2.4 + 2\cdot (D-0.15), &  b_A \geq 6.41\\ 
2.4, &		b_A < 6.41
\end{array}
\right. 
\end{array}
\end{equation}

The ADF steepness parameter $b_A$ is treated as an independent variable during the 
minimization procedure. If, however, 
the core position is far from the dense parts of the 
array, $b_A$ is treated as a fixed parameter. In this case it's value is derived from the 
value of $X_{max}$, obtained in turn from the mean width $\tau_{eff}$ of 
Cherenkov light 
pulses at a distance of $R = 400$\,m. The connection relation  
$\tau_{eff}$ vs. $X_{max}$ and $b_A$ vs. $X_{max}$ has been obtained and 
discussed in \cite{l7}.	

An example of CORSIKA simulated ADF fitted with formula (3) is shown in Fig.\,\ref{fig:6}.
Figure\,\ref{fig:7} presents the result of fitting some hundreds of simulated events for
different zenith angles, different energies and different sorts of nuclei listed above as
the correlation between $b_A$ and the thickness of matter between the EAS maximum and the 
array.   
An example of core reconstruction is presented in Fig.\,\ref{fig:8}. The appropriate functions 
ADF (3) and LDF (see expression (6) below) for this event are presented 
in Fig.\,\ref{fig:9}.

\subsection{Energy reconstruction}

As a measure of energy we use the light flux density 
at a distance $R = 200$\,m -- $Q(200)$. Reconstruction of $Q(200)$ is made by fitting 
the measured values of $Q_i$ with the lateral distribution function (LDF). 
The new expression  used now differs slightly from our previous version
described in \cite{t2009}. 

\begin{equation}
Q(R)=Q(200)\cdot q(R),
\end{equation}

The function $q(R)$ is a combination of three different parametrizations according 
to the distance $R$ to the shower core (in meters):

\begin{equation}
 q(R)=\left\{
\begin{array}{lc}
\!\left(\frac{200}{R_{\mbox{kn}}}\right)^{b_2}\cdot 
\exp\left(\frac{R_{\mbox{kn}}-R}{R_0}\left(1+\frac{3}{R+3}\right)\right),\\
\hspace{25mm} R<R_{\mbox{kn}} \\
\!\left(\frac{200}{R}\right)^{b_2},\hspace{7mm} R_{\mbox{kn}}\leq R\leq
300\,\mbox{m}\\
\!\left(\frac{2}{3}\right)^{b_2} \cdot 
\left(\left(\frac{R}{300}+1\right)/{2}\right)^{-b_Q},\\
\hspace{25mm} R>300\,\mbox{m}\\
\end{array} \right.
\end{equation}

All three variables in equation (3) ($R_0$, $R_{\mbox{kn}}$
and $b_2$), describing the LDF shape in the different ranges of core distance $R$
are related to a single parameter of the LDF shape --
the steepness $b_Q$:

\begin{equation}
\begin{array}{lc}
R_0 = 200\cdot (b_Q - 2.6)^{1.76},\hspace{4mm} \\
R_{\mbox{kn}} = 100 + 38\cdot (b_Q-4.5)^2,\hspace{6mm} b_Q\le 4.5 \\
R_{\mbox{kn}} = 100,\hspace{34mm} b_Q > 4.5 \\
b_2 = 1.97 + 0.12\cdot (b_Q - 3.314)^2
\end{array}
\end{equation}

\begin{figure}[ht]
\includegraphics[width=75mm]{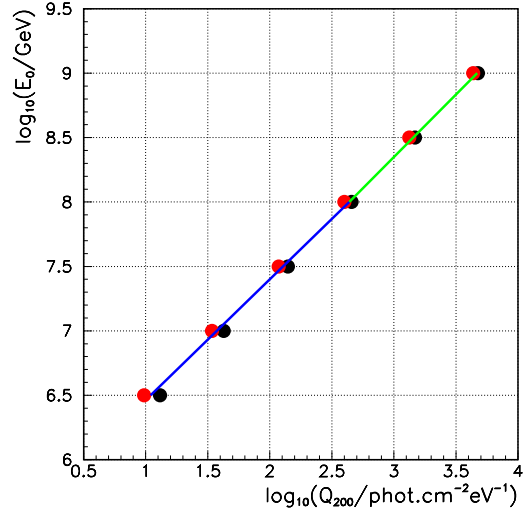} 
\caption{CORSIKA QGSJET-II-04 simulation of $Q(200)$ vs. $E_0$. 
Black points are the mean values of $Q(200)$ for primary protons, red points are the mean 
values of $Q(200)$ for primary iron nuclei.
}
\label{fig:10}
\end{figure}

The connection between the EAS energy $E_0$ and 
$Q(200)$ has been obtained from CORSIKA simulations. 
The result of the simulation using QGSJET-II-04 is shown in Fig.\,\ref{fig:10}. 
Black points are the mean values of $Q(200)$ for primary protons, red points are the mean 
values of $Q(200)$ for primary iron nuclei.
The simulation 
was made for discrete energies  3, 10, 30, 100, 300, 1000, 3000 and 10000 PeV and zenith 
angles $30^{\circ}$ and $45^{\circ}$. 
A simplified primary composition was supposed, consisting of equal 
contributions of protons and iron nuclei ($\ln A = 2.0$), for energies
$\leq 3\cdot 10^{17}$\,eV. The pure proton composition is supposed for the energy $10^{18}$\,eV
in accordance with the results of TA \cite{l15} and PAO \cite{l16} experiments.

A linear fit of the individual simulated points gives
  
\begin{equation}
\log_{10}(E_0) = C_A + (0.940 \pm 0.003)\cdot \log_{10}(Q(200)),    
\end{equation} 

for fitting over the energy range $3\cdot 10^{15} - 10^{17}$\,eV and

\begin{equation}
\log_{10}(E_0) = C_B + (0.951 \pm 0.002)\cdot \log_{10}(Q(200)),    
\end{equation}

for fitting over the energy range $10^{17} - 10^{18}$\,eV. 

The constant $C_A$ is obtained from the procedure of absolute energy calibration
(by the normalization to the reference CR integral intensity) described at the next 
subsection. The constant $C_B$ can be obtained from $C_A$ and the expressions (8) and (9):

\begin{equation}
C_B = (0.951\cdot C_A - 0.01\cdot 17)/0.940,    
\end{equation} 

We have analyzed a possible influence of the uncertainty in composition to the calculation of $E_0$ from
$Q(200)$. Changing the iron percentage
in the composition from \mbox{30 \%} to \mbox{70 \%} changes $\ln A$ from 
1.2 to 2.8 and
the value of the coefficients at the expressions (8) and (9) by less than 
$\pm 0.01$. This is one of the sources of the systematic 
uncertainty in the absolute energy estimation.


\subsection{Absolute energy calibration}

To reconstruct the EAS primary energy from the
Cherenkov light flux measurements one needs to know the
absolute sensitivities of the Cherenkov detectors and the atmosphere 
transparency. To
avoid these problems, the method of normalization of the integral energy
spectrum to a reference spectrum is used.
The reference energy spectrum was measured by the QUEST 
experiment \cite{quest}, \cite{cris_06}.
The procedure of the normalization starts by the precise measurement of the EAS 
(or the primary particle) integral flux for the measured codes (or the relative energy).
To measure the integral flux the number of events exceeding a measured relative code is calculated. 
The integral flux is obtained by dividing this number by the selected effective area, the selected 
effective solid angle and the observation time.      
The so obtained (for each night of Tunka-133 operation) integral 
energy spectrum is compared at the fixed flux level with the reference spectrum point. This comparison 
provides the normalisation constant $C_A$ in expressions (8), used for calculation of the absolute primary 
particle energy of all the events independently of their EAS core position.  

The exact point for normalization is chosen depending on the threshold 
of the experiment, because it has to have the maximal possible statistics for the events, registered with 
the efficiency close to 100\%. It was $3\cdot 10^{15}$\,eV for the Tunka-25 array \cite{t25} 
where different PMTs (QUASAR-370) with about 4 times larger photocathode area were used. 
The normalization energy point for the first 3 years of Tunka-133 operation was 
estimated as $6\cdot 10^{15}$\,eV. About 5 years ago the authorities of the nearest
villages (situated 3 km from the array site) installed  new street lights much 
brighter than the previous ones, so the level of artificial light background became 
much higher. Therefore we used a normalization energy of $10^{16}$\,eV for the last 4 years. 

The systematic uncertainty in the reconstructed primary energy was
estimated in Ref.~\cite{cris_06} to be less than 10\%.
 
This systematic error 
together with the above mentioned uncertainty ($\pm 0.01$) in the coefficients (0.940 and 0.951) in  
formulae\,8 and 9 provides the systematic uncertainty in the primary energy spectrum
which will be shown in Fig.\,\ref{fig:15}.

\subsection{Experimental evaluation of the accuracy of the main EAS parameters}
\label{sec-3}

\begin{figure}[h]
\includegraphics[width=18pc]{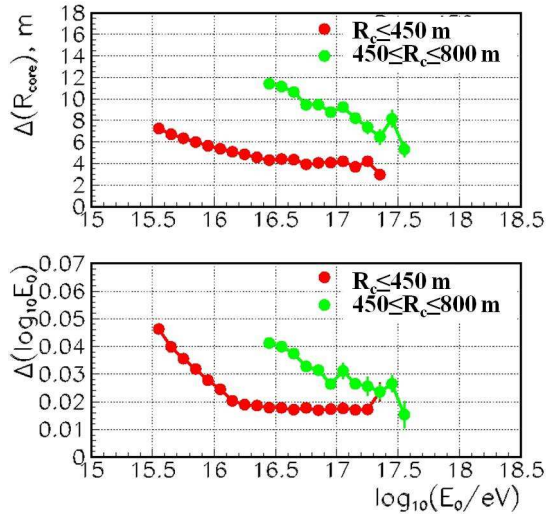}
\caption{\label{fig:11} Upper panel -- errors of core position reconstruction. Lower panel --
relative errors of primary energy reconstruction.}
\end{figure}
\begin{figure}[h]
\includegraphics[width=18pc]{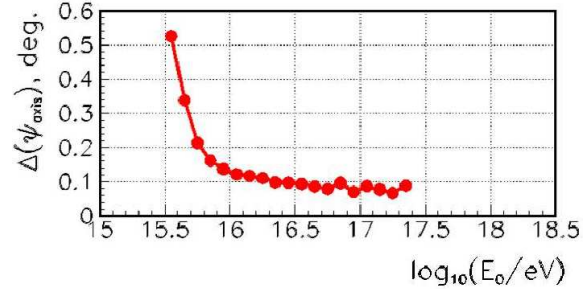}
\caption{\label{fig:12} Error of arrival direction.}
\end{figure}

The accuracy of the reconstructed  shower parameters can be estimated using the well
known chessboard method \cite{chess}. To use this method, the array is divided
into two independent sub-arrays of similar size and configuration. Then the EAS parameters
are derived independently with each of the two sub-arrays. The accuracy of any parameter
is given by the difference between the two reconstructed values divided by a factor of 
$\sqrt 2$ because
they represent two independent experimental determinations of the same shower parameter.
Applying this method to the Tunka-133 data, we composed the first sub-array from detectors with
odd station IDs 
and the second one from detectors with even station IDs. The results are shown 
in Figs\,\ref{fig:11} and\,\ref{fig:12}.

The upper panel of Fig.\,\ref{fig:11} shows the errors of the reconstruction
of the  core position $R_c$. In the central part of the array ($R_c\leq 450$\,m)
where showers with $\log_{10} E_0 \geq 15.5 $ are recorded with almost 100\%
efficiency (see section 5), the error in $R_c$ is less than 8 m. For the outer,
sparsely instrumented region (450\,m $\leq R_c \leq$ 800\,m) a 100\% efficency is
reached only for $\log_{10} E_0 \geq 16.5 $ with an error in $R_c$ almost twice of
that for the inner region. The lower panel of Fig.\,\ref{fig:11} shows the
relative uncertaintly  in the primary energy $E_0$. For the central part of
the array, the error in $\log_{10} E_0 $ decreases  rapidly from 0.05 at 
$\log_{10} E_0 \sim 15.5 $ to 0.02 for $\log_{10} E_0 > 16.0 $. Even for events
with their core in the peripheral regions  and $\log_{10} E_0 > 16.5 $, the
error is smaller than 0.04. These values are much smaller than the binning of
our energy spectra ($\Delta_{bin}\log_{10} E_0 = 0.1) $ avoiding a spectral
deconvolution of the latter. Fig.\,\ref{fig:12} shows the error of the arrival
direction. For $\log_{10} E_0 \geq 16$, it becomes smaller than 0.12$^\circ$.

\section{Energy spectrum}
\label{sec-4}

\subsection{Experimental data}

Tunka-133 operates in clear moonless nights from October to early April.
During the other months, the nights are 
too short for economical operation and the weather conditions are mostly unsatisfactory. 
Here we present the data from 7 seasons. The total operation time with good weather conditions is 
2175\,hrs. The mean trigger
rate is about 2\,Hz. The number of recorded events is $\sim 1.4 \cdot 10^{7}$.

To reconstruct the differential energy spectrum the number of events inside 
each bin of 0.1 in $\log E_0$ is calculated. The differential uncorrected
intensity ($I_{uc}$) is 
obtained by dividing this number by the selected effective area, the selected effective solid angle, 
the observation time and the energy bin width.    
We have selected events with zenith angles $\theta \le 45^{\circ}$.

\begin{figure}[ht]
  \includegraphics[width=75mm]{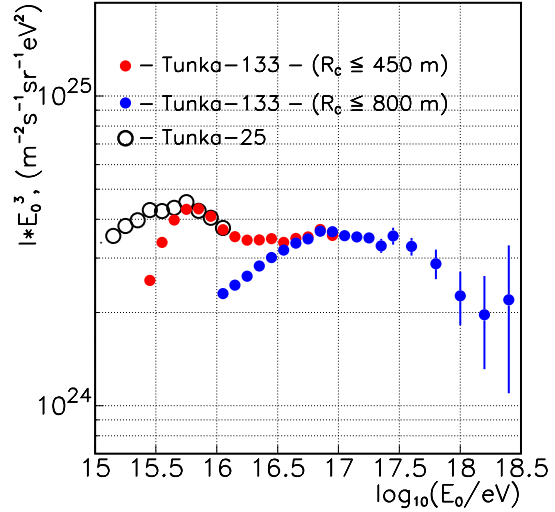}
  \caption{Differential primary cosmic-ray intesity (uncorrected for registration efficiency) 
  I$_{uc}$ for two collecting areas: 
$R_C \le 450$~m and $R_C \le 800$~m -- before combining.}
  \label{fig:13}
 \end{figure}

Fig.\,\ref{fig:13} shows I$_{uc}$ for the very dense array Tunka-25 \cite{t25},
for the dense inner part of Tunka-133 ($R_c <450$\,m) and for full Tunka-133 
($R_c <800$\,m). \mbox{Tunka-25} had a much lower energy threshold than Tunka-133 
(it is $\sim$ 100\% efficient for $\log_{10} E_0 \geq 15$) and can be used to
experimetally derive the Tunka-133 efficiency in the overlapping region.
One sees that for the inner region ($R_c \leq 450$\,m, red dots) full efficiency is
reached at $\log_{10} E_0 \sim 15.7$ while for the full array, i.e. adding the outer
sparsely intrumented regions with 450\,m $\leq R_c \leq$ 80\,m (blue dots) it is reached only at
$\log_{10} E_0 \sim 16.7$. In the outer regions the condition for a shower to be recorded was chosen
to be two or more hit clusters, since for a single-cluster event there is no measurement of the Cherenkov
light flux at a core distance 200 m, whis is used for the energy evaluation.



\begin{figure}[ht]
  \includegraphics[width=75mm]{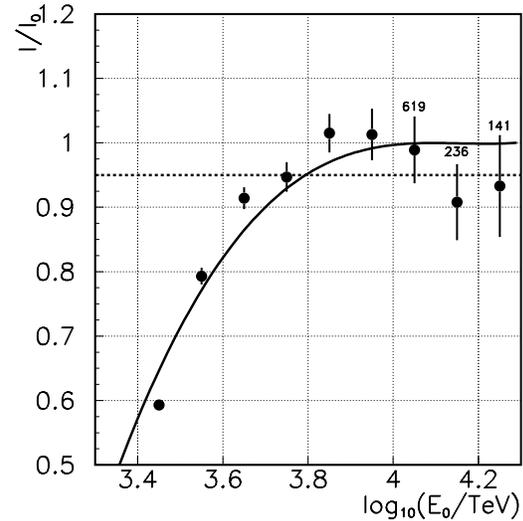}
  \caption{Energy dependence of the EAS registration efficiency in the 
  threshold energy range. Points are the experimental estimation from Tunka-25 
and Tunka-133 data comparison, the solid line is a result of simulation, 
the dotted line is the
efficiency level for the events used for the spectrum reconstruction.}
  \label{fig:14}
 \end{figure}

The Monte-Carlo-calulated efficiency of the inner part of Tunka-133 ($R_c \leq$ 450\,m)
as a function of shower energy is shown in
Figure\,\ref{fig:14} as the ratio of the number of recorded events to the number of generated
events.
At $\log_{10} E_0 \sim 15.35$ the MC calulated efficiency (curve) is $\sim$ 50 \%  and
rises to an almost constant value of 95-100\% for $\log_{10} E_0 \geq 15.8$, in accordance with the 
efficiencies derived from the uncorrected  intensities shown at Figure\,\ref{fig:13} (red
dots). The decreasing efficiency of I$_{uc}$ for  $\log_{10} E_0 < 15.8$ is due to the fact that
below this energy showers are less efficiently recorded even in the inner, densely instumented part.

To 
reconstruct the 
spectrum for $\log_{10} E_0 < 17$ (where we are not limited by statistics) we used only events with
$R_c \leq$ 450\,m, since they have more hits and consequently a better quality than events with the
same energy but $R_c \geq$ 450\,m. For $\log_{10} E_0 > 17$
(where according to Fig.\,\ref{fig:13} also the full array is 100\% efficient) we used all
events with $R_c <800$\,m. For the highest points of the spectrum, with progressively low
statistics, we doubled the bin size from 0.1 to 0.2 in $\log_{10} E_0$. The number of events with 
$R_c \leq$ 800\,m and $\log_{10} E_0 \geq 17$ is 4224.

Due an increasing level of light from artificial sources in a nearby village, we had to raise our
triggering theshold after the first three years. This affects only the first two points of the
spectrum.
For the reconstruction of these two points we therefore use only data from the first 3 years of operation.

The resulting differential energy spectrum is shown in 
Fig.\,\ref{fig:15} together with
the previous spectrum of Tunka-25 \cite{t25}. 
This and the following figures of course show the points with efficiency close to 100\%
according to Fig.\,\ref{fig:13} and  Fig.\,\ref{fig:14}. 

The green contours bracket the possible
systematic errors (described above in the section 4.5) due to the absolute normalisation 
procedure and - to a lesser degree - to the uncertainty 
of the coefficients (0.940 and 0.951) in expressions (8) and (9), connected with the above mentioned
uncertainty in the mass composition. A digital representation of the 
spectrum is presented in Table~\ref{tab-1} together with 
the number of events, statistical and systematic errors.

  \begin{figure}[ht]
  \includegraphics[width=75mm]{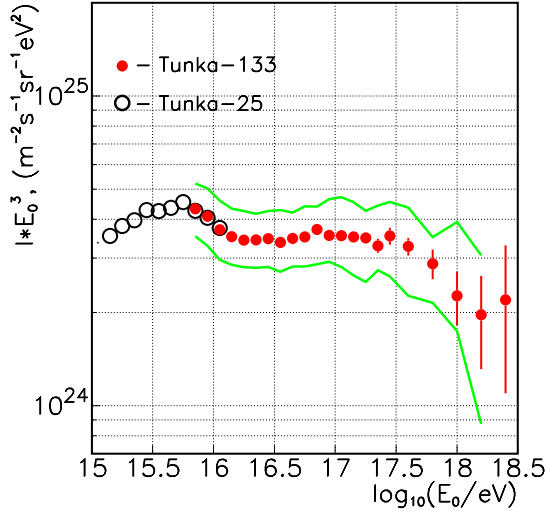}
  \caption{Differential primary cosmic-ray energy spectrum. Green contours bracket 
 the possible systematic errors.}
  \label{fig:15}
 \end{figure}

  \begin{figure}[ht]
  \includegraphics[width=75mm]{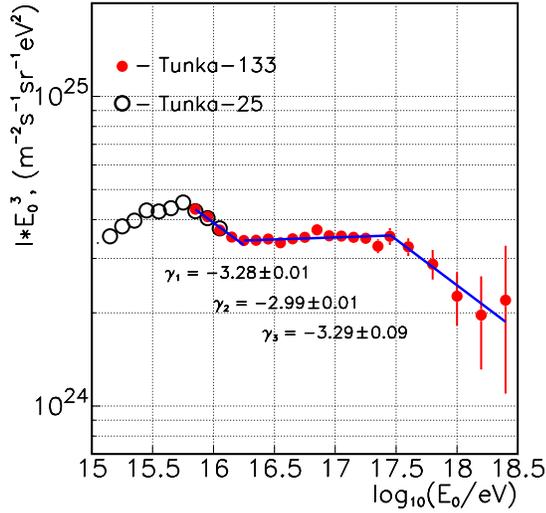}
  \caption{Differential primary cosmic-ray energy spectrum with
  a fit of a doubly broken power law.}
  \label{fig:16}
 \end{figure}
 \begin{table}
\caption{All- particle  energy spectrum.}
\label{tab-1}       

\begin{tabular}{cccr}
$\log(E/eV)$ & $N_{events}$ & & $dI/dE \pm stat\pm sys (m^{-2}sr^{-1}s^{-1} eV^{-1}) $\\\hline
 15.85 & 70199 & & $(1.219 \pm   0.005 \pm 0.261)\times 10{-23}$  \\
 15.95 & 41904 & & $(5.780 \pm   0.028 \pm 1.238)\times 10{-24}$ \\
 16.05 & 53109 & & $(2.621 \pm   0.011 \pm 0.579)\times 10{-24}$ \\
 16.15 & 31788 & & $(1.246 \pm   0.070 \pm 0.262)\times 10{-24}$  \\
 16.25 & 19546 & & $(6.086 \pm   0.044 \pm 1.289)\times 10{-25}$  \\ 
 16.35 & 12363 & & $(3.058 \pm   0.026 \pm 0.614)\times 10{-25}$ \\ 
 16.45 &  7879 & & $(1.548 \pm   0.017 \pm 0.323)\times 10{-25}$ \\ 
 16.55 &  4830 & & $(7.537 \pm   0.011 \pm 1.762)\times 10{-26}$  \\ 
 16.65 &  3142 & & $(3.894 \pm   0.069 \pm 0.779)\times 10{-26}$  \\ 
 16.75 &  2004 & & $(1.973 \pm   0.044 \pm 0.446)\times 10{-26}$ \\ 
 16.85 &  1337 & & $(1.046 \pm   0.029 \pm 0.216)\times 10{-26}$ \\ 
 16.95 &   806 & & $(5.007 \pm   0.176 \pm 1.224)\times 10{-27}$ \\ 
 17.05 &  1604 & & $(2.504 \pm   0.063 \pm 0.682)\times 10{-27}$ \\ 
 17.15 &  1004 & & $(1.245 \pm   0.039 \pm 0.338)\times 10{-27}$\\ 
 17.25 &   629 & & $(6.197 \pm   0.247 \pm 1.616)\times 10{-28}$ \\ 
 17.35 &   374 & & $(2.927 \pm   0.151 \pm 0.696)\times 10{-28}$ \\ 
 17.45 &   254 & & $(1.579 \pm   0.099 \pm 0.423)\times 10{-28}$ \\ 
 17.60 &   237 & & $(5.180 \pm   0.336 \pm 1.639)\times 10{-29}$ \\ 
 17.80 &    83 & & $(1.144 \pm   0.127 \pm 0.303)\times 10{-29}$ \\ 
 18.00 &    26 & & $(2.262 \pm   0.444 \pm 1.000)\times 10{-30}$ \\ 
 18.20 &     9 & & $(4.941 \pm   1.647 \pm 1.647)\times 10{-31}$ \\ 
 18.40 &     4 & & $(1.386 \pm   0.693 \pm 0.314)\times 10{-31}$  \\\hline   
\end{tabular}

\vspace*{1cm}  
\end{table}

  \begin{figure}[!t]
  \includegraphics[width=75mm]{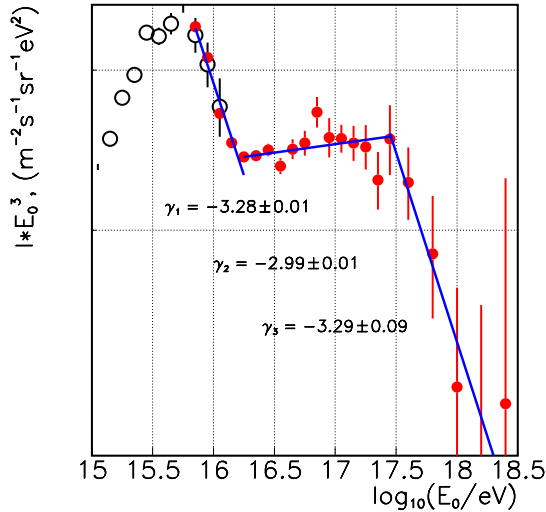}
  \caption{Representation of the energy spectrum with an expanded scale of the y-axis.}
  \label{fig:17}
 \end{figure}

\subsection{Features in the energy spectrum}

The spectrum of Tunka-133
shows a number of features, that is deviations from a single power law. 
A power law applies only for spectral regions extending over less than half an order of magnitude.
A fit with a doubly broken power law is shown in Fig.\,\ref{fig:16}. 
At an energy of 
about $2\cdot 10^{16}$ eV, the power law index changes from 
$\gamma = 3.28\pm 0.01$ to $\gamma = 2.99\pm 0.01$. This feature was first 
observed by the KASCADE-Grande experiment \cite{l4}.
The index can well be used
up to $E_0 = 3\cdot 10^{17}$ eV. But the statistical probability of a single 
power law representation
in this energy range is rather small (about 3\%). The main reason of this is 
one point at $\sim 7\cdot 10^{16}$~eV, 
which is higher than the power law by about 2 standard 
deviations. This can be seen better if  
the y-axis is expanded as in  Fig.\,\ref{fig:17}. 
A similar feature at the same energy was observed in the experiment GAMMA 
\cite{l8}.

The spectrum becomes much steeper 
with $\gamma = 3.29\pm 0.09$ above  $3\cdot 10^{17}$~eV (the second "knee"). 

\subsection{ Discussion}

\begin{figure}[!t]
  \includegraphics[width=75mm]{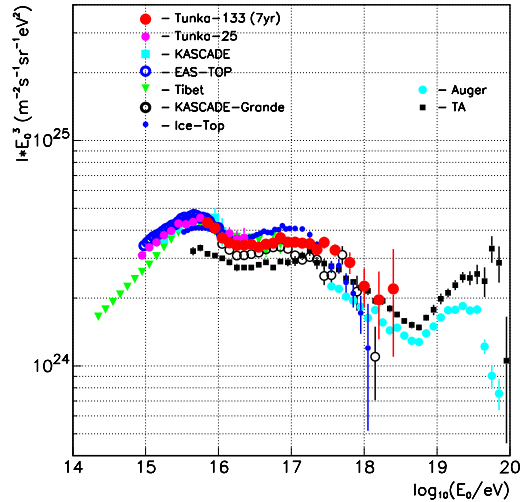}
  \caption{Comparison of energy spectra obtained at the Tunka site 
to other experimental results.}
  \label{fig:18}
 \end{figure}

In Fig.\,\ref{fig:18} the spectrum is compared to results from other experiments.
The spectra of  all the experiments shown in Fig.\,\ref{fig:18} -- KASCADE \cite{l9}, 
EAS-TOP \cite{l5},
Tibet \cite{l10}, IceTop \cite{l12} -- are practically indistinguishable 
at the energy of the first (classical) knee. 

There is  agreement between the spectra of Tunka-133,
KASCADE-Grande \cite{l4} and  IceTop \cite{l12}
in the intermediate energy range $10^{16} - 10^{17}$ eV. 
We note that the difference between Tunka-133 and KASCADE-Grande
spectra at $E_0$ about $10^{17}$ eV can be eliminated by an energy correction
of only 4\%. The energy shift between Tunka-133 data and IceTop data
is slightly larger (about 7\%).
However, these shifts
are smaller than the systematic errors of energy reconstruction in these experiments.
The surprisingly good agreement among the spectra was emphasized  
in \cite{haungs15} and \cite{rex16}. 

The  paper \cite{rex16} is 
devoted to the comparison of energy scales of Tunka-133 and KASCADE-Grande. 
The spectra were not only compared w.r.t. their form, but the absolute scales of both experiments
have been compared via the radio extensions of these experiments named 
Tunka-Rex and LOPES.
The radio detectors have been independently and accurately calibrated. The estimation of the absolute
energy calibration uncertainty is less than 10\% \cite{rex16}.

For highest energies the Tunka-133 data are in agreement with data from the Telescope Array (TA) \cite{l15} 
and the Pierre  Auger Observatory \cite{l16}, but naturally the  
Tunka-133 statistics is rather low above 
$10^{18}$\,eV.

To get more statistics at this energy range, the scintillation counter array Tunka-Grande is
presently being  constructed at the Tunka-133 site\cite{grande17}. The total data taking time will be 20 times 
higher than for the Cherenkov light array. The scintillation detectors will provide information on
the number of electrons and muons in the EAS.

\section{Conclusion}
\label{sec-6}

1. The primary CR energy spectrum in the range of $6\cdot 10^{15} - 10^{18}$ eV 
has a number of features: the spectrum 
becomes harder (the index changes from $\gamma = 3.28\pm 0.01$ 
to $\gamma = 2.99\pm 0.01$) at $E_0 = 2\cdot 10^{16}$ eV and 
steeper ($\gamma = 3.29\pm 0.09$) at $E_0 = 3\cdot 10^{17}$ eV.

2. In the energy range of $10^{16} - 10^{17}$ eV, the observed  spectrum 
is consistent with 
spectra of KASCADE-Grande \cite{l4} and IceTop \cite{l12}.

3. Beyond the energy of $10^{17}$ eV, the Tunka-133 spectrum is consistent with 
those from the Telescope Array \cite{l15} and the Pierre  Auger Observatory \cite{l16}.

\section{Acknowledgment}
\label{acknow}
This work is supported by the Russian Federation Ministry of Science and High Education 
(agreement: 075-15-2019-1631), the Russian Science Foundation (grant 19-72-20067 (section 5),
grant 19-72-20230 (section 4)), the Russian Foundation for Basic Research
(grants 19-52-44002, 16-29-13035).



\end{document}